\pdfoutput=1

\documentclass[final,prl,aps,twocolumn,preprintnumbers,amssymb,nobibnotes,nofootinbib,superscriptaddress,10pt]{revtex4-1}

\usepackage{hyperref,graphicx,array,multirow,color,amsmath,mathrsfs,titlesec,shuffle,slashed}

\titleformat{\section}{\centering\normalsize\normalfont\bf}{\thesection}{1em}{}

\hypersetup{pdftitle={},pdfcreator={},linkcolor=[rgb]{0.15,0.35,0.75},colorlinks=true,citecolor=[rgb]{0.675,0,0.2},urlcolor=[rgb]{0.15,0.35,0.65}}

\newcommand{\smallminus}{{\rm\rule[2.4pt]{6pt}{0.65pt}}}
\newcommand{\smallplus}{\hspace{0.5pt}\text{{\small+}}\hspace{-0.5pt}}

\newcommand{\ab}[1]{\langle #1 \rangle}
\newcommand{\aab}[1]{\langle\hspace{-2.6pt}\langle #1 \rangle\hspace{-2.6pt}\rangle}
\newcommand{\NeqFour}{\mathcal{N}{=}4\,\text{SYM}}
\newcommand{\MHVbar}{\overline{\text{MHV}}}

\newcommand{\bigger}[1]{\raisebox{-0.95pt}{\scalebox{1.25}{$#1$}}}

\begin{document}

\preprint{}

\title{Positive Geometries for One-Loop Chiral Octagons}
\author{Enrico Herrmann}
\affiliation{SLAC National Accelerator Laboratory, Stanford University, Stanford, CA 94039, USA}

\author{Cameron Langer}
\affiliation{Center for Quantum Mathematics and Physics (QMAP), University of California, Davis, CA, USA}

\author{Jaroslav Trnka}
\affiliation{Center for Quantum Mathematics and Physics (QMAP), University of California, Davis, CA, USA}

\author{Minshan Zheng}
\affiliation{Center for Quantum Mathematics and Physics (QMAP), University of California, Davis, CA, USA}

\begin{abstract}
%
Inspired by the topological sign-flip definition of the Amplituhedron, we introduce similar, but distinct, positive geometries relevant for one-loop scattering amplitudes in planar $\mathcal{N}=4$ super Yang-Mills theory. The simplest geometries are those with the maximal number of sign flips, and turn out to be associated with chiral octagons previously studied in the context of infrared (IR) finite, pure and dual conformal invariant local integrals. Our result bridges two different themes of the modern amplitudes program: positive geometry and Feynman integrals.
\end{abstract}

\maketitle

\section{Introduction}
\label{sec:intro}
\vspace{-.4cm}
%
Recently, a number of intriguing connections between \emph{positive geometries} \cite{Arkani-Hamed:2017tmz} and scattering amplitudes in various quantum field theories have surfaced. The primary example of a positive geometry is the Amplituhedron \cite{Arkani-Hamed:2013jha,Arkani-Hamed:2017vfh}, which generalizes convex polygons to Grassmannians. Associated to these geometries are differential forms with logarithmic singularities on all their boundaries. From a physics perspective, these forms reproduce all tree-level amplitudes and the all-loop integrand in planar maximally supersymmetric Yang-Mills theory ($\NeqFour$ \cite{sYM1,sYM2}). Here, planarity refers to the 't Hooft limit~\cite{tHooft:1973alw}, $N_c{\to}\infty$, of $\textrm{SU}(N_c)$ gauge theory\!. For more recent explorations of the Amplituhedron, see e.g.~\cite{Arkani-Hamed:2014dca,Dennen:2016mdk,Ferro:2015grk,An:2017tbf,Arkani-Hamed:2018rsk,Karp:2017ouj,Galashin:2018fri,Damgaard:2019ztj,Lukowski:2020dpn}.

Another positive geometry is the Associahedron \cite{Arkani-Hamed:2017mur} which plays a role in biadjoint $\phi^3$ theory. Ongoing efforts also attempt to extend a similar geometric framework to amplitudes in other field and string \cite{Arkani-Hamed:2019mrd,He:2020ray} theories, and to explore the origin and consequences of this connection.

In this letter, we focus on the positive geometry of one-loop amplitudes in planar $\NeqFour$ (computed first in \cite{Bern:1994zx,Bern:1994cg}), motivated by the Amplituhedron construction. We show that for certain classes of positive geometries, their associated logarithmic forms are given by chiral one-loop integrals. For external kinematics relevant to MHV amplitudes, the most general geometry corresponds to \emph{chiral octagons}. These objects have been defined in \cite{ArkaniHamed:2010gh} as a basis of one-loop dual conformal invariant integrals with special infrared (IR) properties. In particular, this basis is naturally divided into IR divergent integrals, IR finite integrals, and those which integrate to zero.

We explicitly show that each chiral octagon is associated with a single `local positive geometry' that has similarities with the Amplituhedron, but differs in the choice of topological sign-flip conditions (introduced below). Importantly, while the Amplituhedron geometry increases in complexity with the number of external particles (to capture the complexity of higher point amplitudes), the positive geometries for chiral octagons remain unchanged beyond eight points. Our result constitutes the first example of a connection between positive geometry and Feynman integrals, and we devote a longer companion paper \cite{LocTriangToAppear} for more detailed analysis.

This letter is organized as follows. In section 2, we define the Amplituhedron geometry at one-loop using the sign-flip definition of \cite{Arkani-Hamed:2017vfh}. In section 3, we explore the general one-loop sign-flip regions for MHV kinematics, and show that there is an upper bound on the number of sign flips in certain kinematic quantities. Surprisingly, the maximal sign-flip regions correspond to chiral octagons which we review in section 4. In section 5, we comment on the positive geometries for other integrals which can be deduced from their $d\log$ representations. In section 6, we study the spaces for non-MHV kinematics and show that the logarithmic forms are chiral integrals with non-unit leading singularities which also appeared in \cite{ArkaniHamed:2010gh} in the context of the one-loop ratio function. We end with some conclusive remarks and future directions.

\vspace{-.5cm}
\section{One-Loop Amplituhedron}
\label{sec:amplituhedron}
\vspace{-.4cm}

Before describing our novel local positive geometries, we briefly discuss the well-known Amplituhedron construction in order to point out similarities and differences between these spaces. Our main framework will be the topological definiton of the Amplituhedron in terms of certain sign-flip conditions \cite{Arkani-Hamed:2017vfh} on kinematic invariants. Due to the special symmetries of planar $\NeqFour$, momentum-twistor space \cite{Hodges:2009hk} is the natural kinematic setting to describe scattering amplitudes in this theory. The external kinematics for massless $n$-particle scattering amplitudes are encoded in $n$ four-vectors $Z^I_a$, $I{\in}\{1,{\ldots},4\}$, $a{\in}\{1,{\ldots},n\}$ defined up to little-group transformations, $Z_a \sim t_a Z_a $. The (dual) spacetime symmetries of planar $\NeqFour$ act linearly on the $Z_a$ via $\textrm{SL}(4)$ transformations, and the invariants are given by contractions of four momentum twistors with the four-dimensional Levi-Civita tensor $\ab{ijkl} {:=} \epsilon_{IJKL}Z^I_iZ^J_jZ^K_kZ^L_l$.

Loop momenta are encoded via lines in twistor space. Each line is defined by the linear span of two representative points, so that we may associate $\ell \leftrightarrow (AB)$. Loop-dependent quantities involve contractions of $(AB)$ with bi-twistors $X^{IJ}$. In the following, we encounter various four-brackets of the form $\ab{ABX}$, although the most important correspond to inverse propagators of the form $\ab{ABii{+}1}$. For further details on the twistor correspondence, see e.g.~\cite{Hodges:2009hk,Mason:2009qx,Mason:2010yk,Arkani-Hamed:2013kca}. Throughout, the notation $\aab{X}{:=}\ab{ABX}$ is used for brevity. With the basic kinematic objects in hand, the topological definition \cite{Arkani-Hamed:2017vfh} of the $L$-loop $n$-point, $\text{N}^k\text{MHV}$ amplituhedron $\mathcal{A}^{(n,k,L)}$ is:
\begin{align}
\label{eq:AnkL_generic_def}
\begin{split}
 & \bullet \ab{ii{+}1jj{+}1} >0\,, \\
 & \bullet  \text{sequence } \{\ab{abb{+}1i}\}_{i\neq a,b,b{+}1} \text{ has $k$ sign flips}, 
 \\[3pt]
 \hline 
 \\[-12pt]
 & \bullet \aab{ii{+}1}>0, \\
 & \bullet  \text{sequence } \{\aab{1i}\}_{i\neq 1} \text{ has $k{+}2$ sign flips,}\\
 & \bullet \ab{(AB)_i(AB)_j}>0 \text{ for $i\neq j=1,\ldots,L\geq2$.}
\end{split}
\end{align}
At one loop there is an equivalent definition of the MHV ($k=0$) Amplituhedron:
\begin{align}
\label{eq:MHV_A_def}
\text{MHV def.: }
\begin{split}
 & \bullet \ab{ijkl} >0\,, \text{for } i{<}j{<}k{<}l, \\
 & \bullet \aab{\overline{ij}}>0, \text{ for all $i{<}j$,}
\end{split}
\end{align}
where $(\overline{ij})$ denotes the line defined as the intersection of two planes  $(\overline{ij}){:=}(i{-}1ii{+}1){\cap}(j{-}1jj{+}1)$. In momentum-twistor space, parity is implemented via the duality between points and planes: $\mathbb{P}:\ a \leftrightarrow \overline{a}:= (a{-}1 a a{+}1)$, and suggests another `$\MHVbar$' space,
%
\begin{align}
\label{eq:MHVb_A_def}
\text{`$\MHVbar$' def.: }
\begin{split}
 & \bullet \ab{ijkl} >0\,, \text{for } i{<}j{<}k{<}l, \\
 & \bullet \aab{ij}>0, \text{ for all $i{<}j$.}
\end{split}
\end{align}
Note that the space in eq.~(\ref{eq:MHVb_A_def}) is \emph{not} the `actual' $\MHVbar$ space defined by setting $k{=}n{-}4$ in eq.~(\ref{eq:AnkL_generic_def}) \cite{Langer:2019iuo}. Both spaces differ by the conditions imposed on the external data, but the associated loop-dependent differential forms are trivially related by stripping the overall tree-level amplitude factor.

Since $\aab{ii{+}1}$ and its dual, $\aab{\overline{ii{+}1}}$, are related as $\aab{\overline{ii{+}1}} = \aab{ii{+}1} \ab{i{-}1 i i{+}1 i{+}2}$, the chiral one-loop MHV Amplituhedron can be viewed as a subspace of a larger \emph{achiral} space defined by a smaller set of inequalities,
\begin{equation}
    S^{(0)}: \aab{ii{+}1}>0, \text{ for all $i{<}j$.}
    \label{sf0}
\end{equation}
In order to recover the MHV Amplituhedron, which we denote $S^{(0),m}$, from eq.~(\ref{sf0}), we ought to impose $\aab{\overline{ij}}{>}0\,, \text{for }i{<}j$. While these extra conditions are necessary to define the chiral subspace, none of them correspond to physical boundaries of the geometry. The question of whether or not inequalities are accessible as geometric boundaries is quite subtle, and we discuss these aspects at greater length in ref.~\cite{LocTriangToAppear}. 

The complement of $S^{(0),m}$ in $S^{(0)}$, denoted as $S^{(0),\overline{m}}$, is also a space with only physical boundaries. Its logarithmic form corresponds to the $\MHVbar$ one-loop amplitude. From eq.~(\ref{eq:MHVb_A_def}), it follows that we can carve out this chiral subspace from $S^{(0)}$ by imposing the conditions $\aab{ij}>0$.

\vspace{-.5cm}
\section{One-Loop Positive Geometries from Sign Flips}
\label{sec:1L_pos_geometry}
\vspace{-.4cm}

Having discussed $S^{(0)}$ in eq.~(\ref{sf0}) where all $\aab{ii{+}1}$ are positive, it is natural to wonder about geometric spaces and canonical forms where a subset of the $\aab{ii{+}1}$ inequalities change sign.  In this section, we restrict our analysis to the one-loop case with the MHV external kinematics of eq.~(\ref{eq:MHV_A_def}). Starting from the achiral space in eq.~(\ref{sf0}), we flip the signs of some $\aab{ii{+}1}$ brackets from positive to negative. Surprisingly, we find that `most' of these spaces are empty for general $n$, i.e. the inequalities are mutually inconsistent; only a very limited number of cases are allowed. We can split these cases into three categories based on the number of sign flips in the sequence (note: this is different than (\ref{eq:AnkL_generic_def})):
\begin{equation}
    \mathcal{P}=\{\aab{12}, \aab{23},\dots, \aab{n{-1}n},\aab{1n}\}\,.
    \label{seq}
\end{equation}
If there are no sign flips in ${\cal P}$, all brackets can be chosen positive; this is nothing but the original $S^{(0)}$ space. For obvious reasons, we call $S^{(0)}$ a \emph{sign-flip-zero} space.

The other non-empty spaces have either two or four sign flips in ${\cal P}$ and are denoted by $S^{(2)}$ and $S^{(4)}$, respectively. We represent these sequences graphically as circles where the signs of $\aab{ii{+}1}$ are marked on the periphery.
\begin{equation}
\label{eq:MHV_achiral_sf_spaces}
\begin{tabular}{|c|c|c|}
\hline
$S^{(0)}$ & $S^{(2)}_{ij}$ & $S^{(4)}_{ijkl}$ \\
\hline
\!\includegraphics[scale=.4]{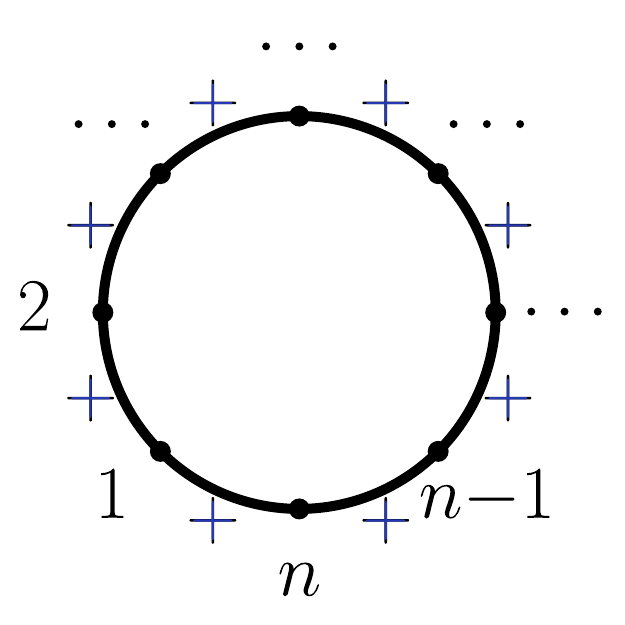}\!
&
\!\includegraphics[scale=.4]{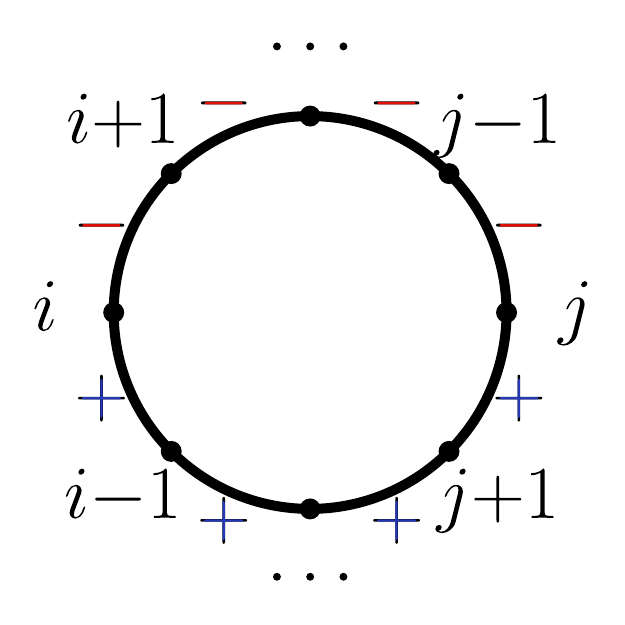}\!
&
\!\includegraphics[scale=.3]{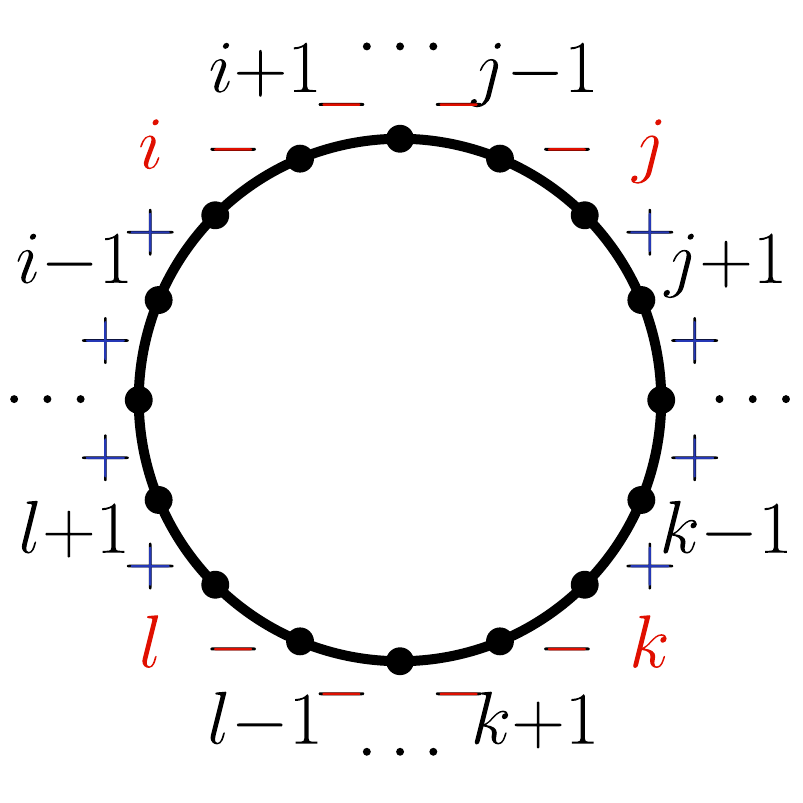}\!
\\[-5pt]
\hline
\end{tabular}
\end{equation}
For the sign-flip two and four spaces, we also indicate the positions where the corresponding sign flips occur. Interestingly, for MHV external kinematics, any space with more than four sign flips is \emph{empty}. 


As mentioned below eq.~(\ref{sf0}), the achiral sign-flip-zero space $S^{(0)}$ can be cut into two chiral components by imposing $n{-}3$ additional inequalities. The two resulting subspaces $S^{(0),m}$ and $S^{(0),\overline{m}}$, are relevant for MHV and $\MHVbar$ amplitudes respectively.

Similarly, we can cut $S^{(2)}$ and $S^{(4)}$ into two chiral components, but surprisingly (in both cases) this can be done by imposing a \emph{single} additional condition.
\begin{align}
\label{eq:sf2_chiral_comps_fully_dressed}
S^{(2),\pm}_{ij} & =
\raisebox{-35pt}{\includegraphics[scale=.47]{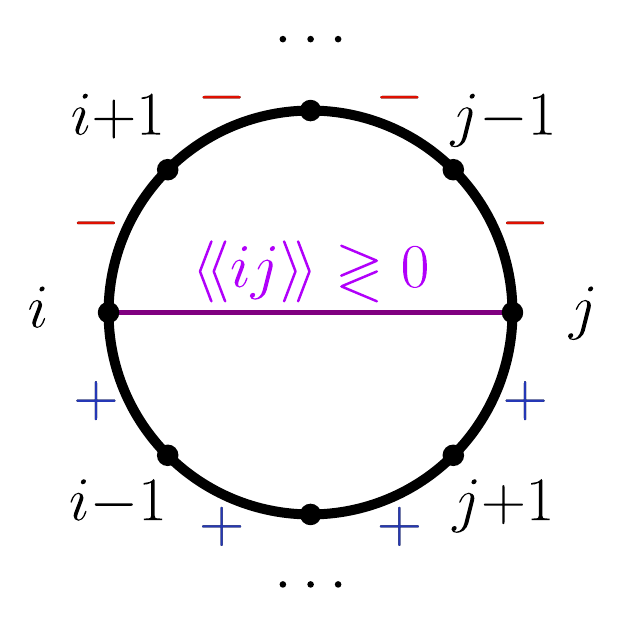}}\,, 
&& \text{for } i<j\,.
\\[-8pt]
S^{(4),\pm}_{ijkl}
& =\raisebox{-38pt}{\includegraphics[scale=.35]{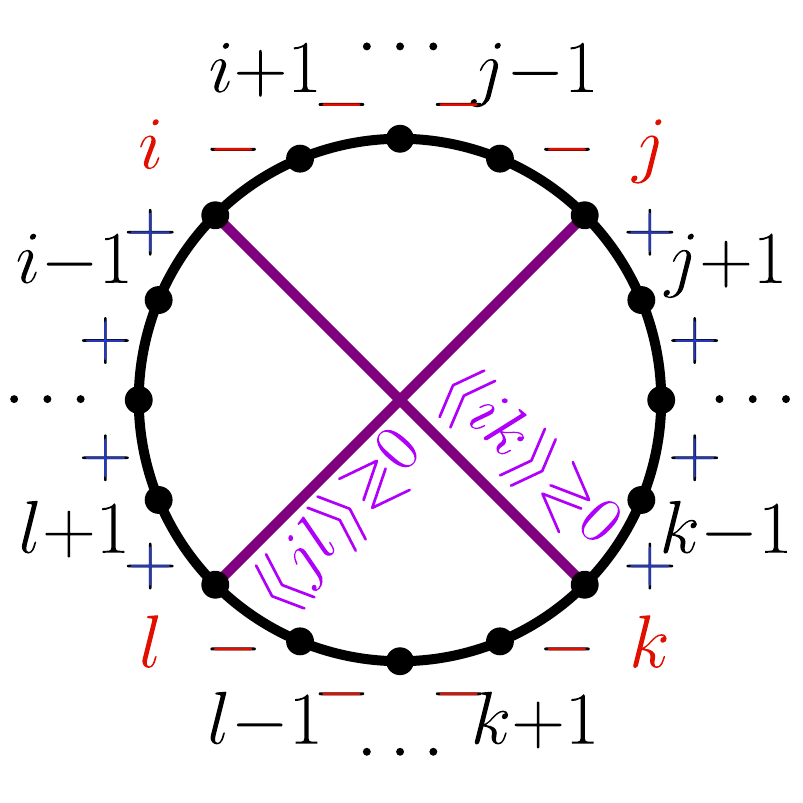}}\,, 
&& \text{for } i<j<k<l\,.
\label{eq:sf4_chiral_comps_fully_dressed}
\end{align}
For $S^{(4),\pm}_{ijkl}$ we indicated two conditions, but either of the diagonal brackets $\aab{ik}$ or $\aab{jl}$ suffices to chiralize the space. The fact that a single condition chiralizes the higher sign-flip spaces suggests that increasing the number of sign flips in ${\cal P}$ leads to simpler positive geometries. We will see this explicitly in the canonical forms for these spaces computed in the following section. The forms for the chiral minimal sign-flip spaces, $\Omega^{(0)}_{\text{MHV},\MHVbar}$, are the full $n$-point MHV and $\MHVbar$ integrands, respectively. These obviously grow in complexity for higher $n$. In contrast, for the maximal sign-flip spaces, $S^{(4),\pm}_{ijkl}$, each form is a single chiral integral. For general $n$, these turn out to be the chiral octagons, previously discussed in \cite{ArkaniHamed:2010gh}. 

As emphasized below eq.~(\ref{eq:MHV_achiral_sf_spaces}), if there are six or more sign flips in the sequence $\mathcal{P}$ the positivity conditions are so restrictive that the space is empty; hence, the corresponding form identically vanishes. Although the differential forms for the sign-flip-two positive geometries are considerably simpler than those of the sign-flip-zero (amplitudes) case, their complexity does grow (mildly) with increasing $n$. Thus, we postpone a full exploration of the sign-flip-two spaces and their forms to \cite{LocTriangToAppear}, while this letter is dedicated to the `maximal' sign-flip-four spaces yielding the simplest local integrals.

\vspace{-.5cm}
\section{Maximal sign flips and chiral octagons}
\label{sec:chiral}
\vspace{-.4cm}
%

As just discussed, the maximal sign-flip-four spaces are simplest, which is why we now focus on these geometries and construct their differential forms. Before jumping to the general result, we give some lower-point examples: 
\begin{align}
\label{eq:sf4_chiral_low_n_egs}
\begin{split}
\hspace{-1cm}
\raisebox{-36pt}{\includegraphics[scale=.42]{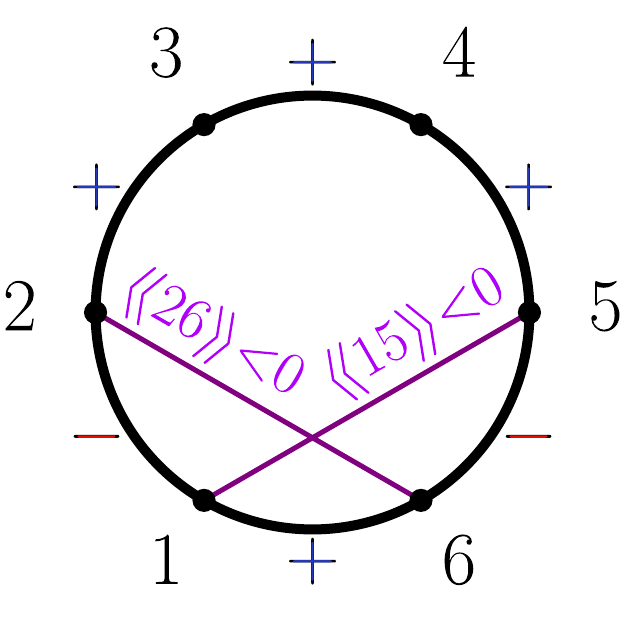}} 
&\leftrightarrow
\frac{d^4\mu\,  \aab{\overline{25}}\ab{1256}}
     {\aab{12}\aab{23}\aab{45}\aab{56}\aab{16}}
{\equiv}
\hspace{-.3cm}
\raisebox{-25pt}{\includegraphics[scale=.33]{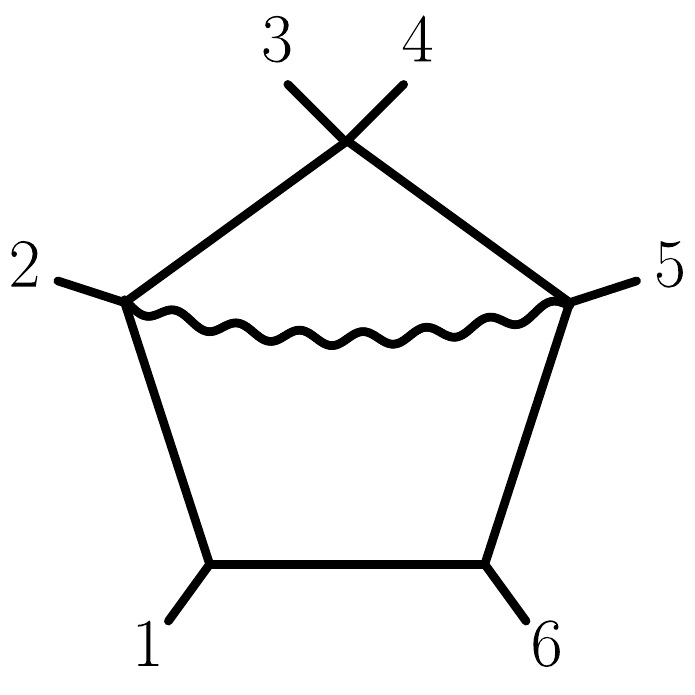}} 
\hspace{-1cm}
\\[-15pt]
\hspace{-1cm}
   \raisebox{-36pt}{\includegraphics[scale=.4]{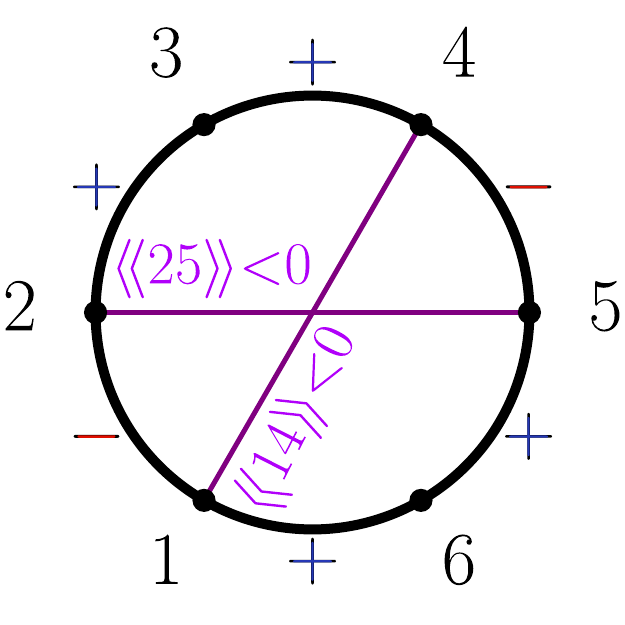}}
   &\leftrightarrow
   \frac{d^4\mu\, \aab{\overline{24}}\aab{\overline{51}}}
        {\aab{12}\aab{23}\cdots\aab{16}}
   {\equiv} 
   \hspace{.4cm}
   \raisebox{-43pt}{
   \includegraphics[scale=.35]{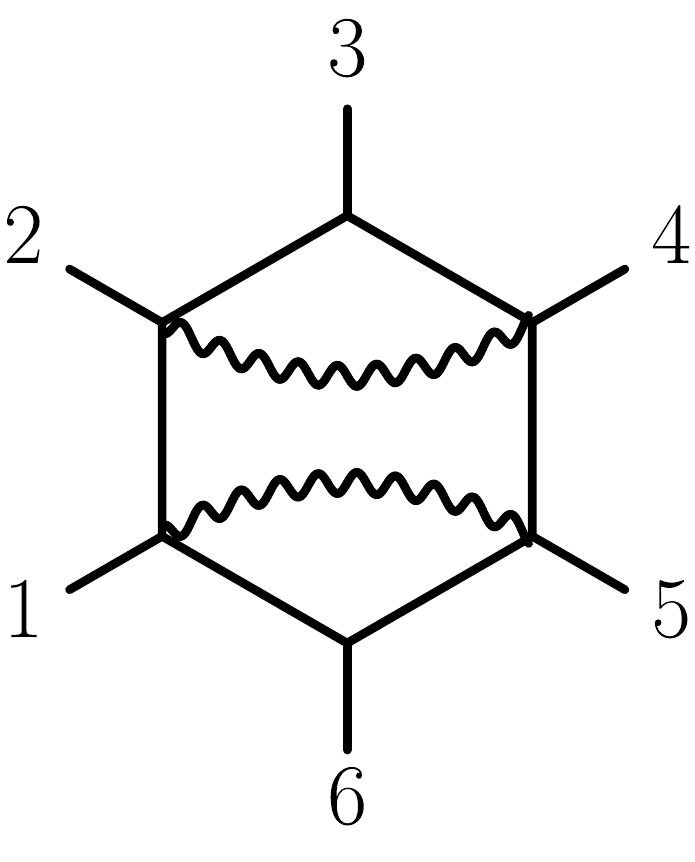}} 
\hspace{-1cm}
\end{split}
\end{align}   
where the measure is $d^4\mu=\ab{ABd^2A}\ab{ABd^2B}$. In the general $n$-point case, the differential form corresponds to octagons with very special chiral numerators 
\begin{align}
\label{eq:chiral_octagon_sf4_connection}
   \raisebox{-39pt}{\includegraphics[scale=.4]{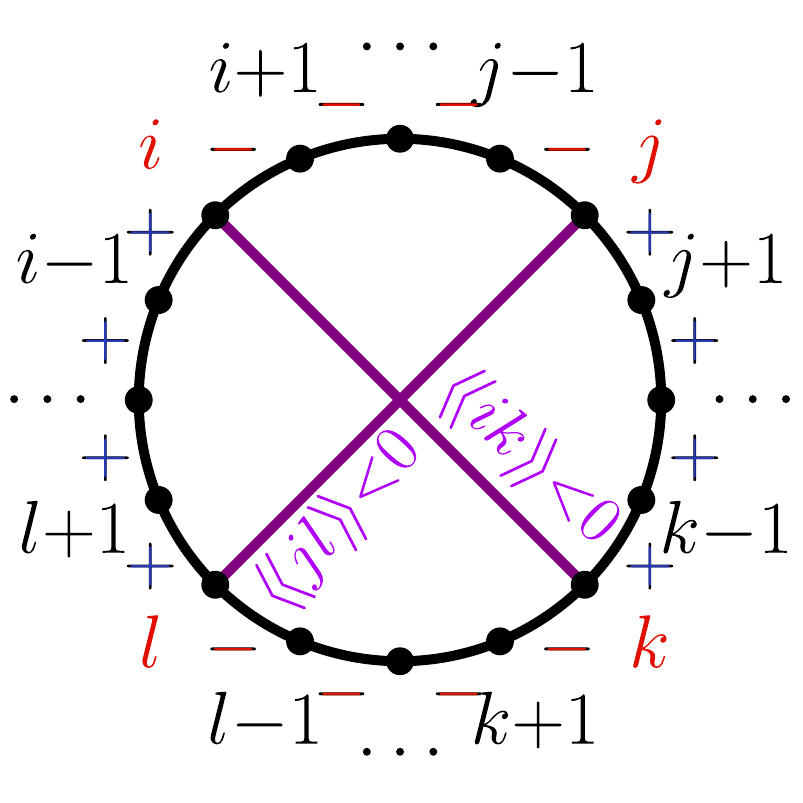}}
   \hspace{.2cm}
   &\leftrightarrow
   \raisebox{-35pt}{
    \includegraphics[scale=.4,angle=0]{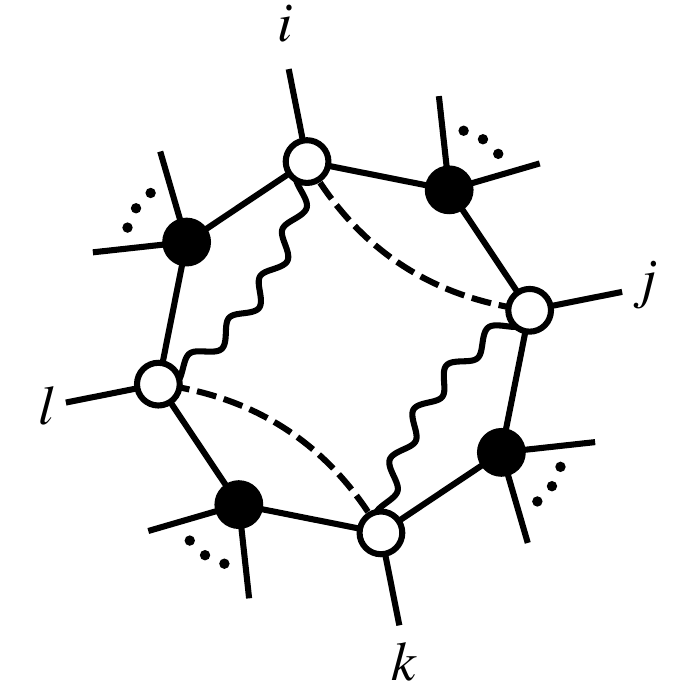}} 
\end{align}
The wavy and dashed lines denote the special numerator factors $\aab{\overline{jk}}{=}\ab{AB(j{-}1jj{+}1){\cap} (k{-}1kk{+}1)}$ and $\aab{ij} {=} \ab{ABij}$, respectively \cite{ArkaniHamed:2010gh}. In eqs.~(\ref{eq:sf4_chiral_low_n_egs}) and (\ref{eq:chiral_octagon_sf4_connection}) we have only displayed one chirality. The opposite chirality is obtained by switching the signs of the non-adjacent $\aab{ij}$ brackets in the definition of the sign-flip-four spaces, as well as interchanging wavy and dashed numerators in the associated integrals. 

It is a highly non-trivial statement that the sign-flip-four geometries have only (up to) eight codimension-one boundaries $\aab{ii{+}1}{=}0$. This simplicity is in stark contrast to the sign-flip-zero and two spaces where all $n$ such boundaries are present. 

Amusingly, the chiral octagons appeared in a previous study of local loop integrands for one-loop scattering amplitudes \cite{ArkaniHamed:2010gh}, completely unrelated to the geometric sign-flip spaces that we investigate in the present letter. In \cite{ArkaniHamed:2010gh}, it was suggested that \emph{chiral octagons} form a natural, and practically useful, basis of one-loop integrands for scattering amplitudes in planar $\mathcal{N}=4$ SYM with many desirable features. In the generic case, we have
\vskip -.6cm
\textcolor{white}{.}
\begin{align}
\label{eq:chiral_octagon_general}
    \hspace{1cm}
    \Omega_{ijkl}  :=&
    \raisebox{-38pt}{
    \includegraphics[scale=.4,angle=0]{./figures/general_octagon_integral}}
    \text{ for } i{<}j{<}k{<}l \\[+2pt]
    & 
    \hspace{-2.3cm}
    {=}
    \frac{d^4\mu\, \aab{ij}\aab{\overline{jk}}\aab{kl}\aab{\overline{li}}}
    {\aab{i{\smallminus}1 i}\!
    \aab{ii{\smallplus}1}\!
    \aab{j{\smallminus}1 j}\!
    \aab{jj{\smallplus}1}\!
    \aab{k{\smallminus}1 k}\!
    \aab{kk{\smallplus}1}\!
    \aab{l{\smallminus}1 l}\!
    \aab{ll{+}1}} \nonumber
\end{align}
These integrands are IR finite and all have unit leading singularities, i.e. all codimension-four residues are either $\pm 1$ or $0$. Spacetime parity has a natural action on these integrands; $\mathbb{P}: \Omega_{ijkl} \mapsto \Omega_{jkli}$, so that one can define parity even and odd combinations of chiral octagons. 
\begin{align}
\label{eq:odd_octagon}
    \Omega^{\text{e/o}}_{ijkl} \equiv 
    \Omega_{ijkl} \pm \Omega_{jkli}
\end{align}
For special leg configurations, the general octagons degenerate into simpler topologies with fewer propagators, 
\vskip -.6cm
\textcolor{white}{.}
\begin{align}
\label{eq:octagon_degenerations}
\hspace{-1cm}
\raisebox{-1.05cm}{
    \includegraphics[scale=.33]{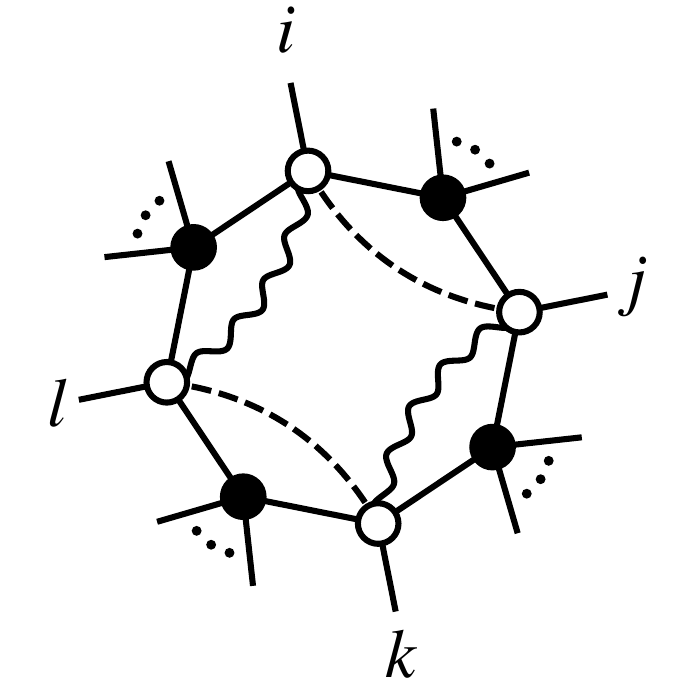}}
    \hspace{-.3cm}
    \to
    \hspace{-.4cm}
\raisebox{-1.1cm}{
    \includegraphics[scale=.33]{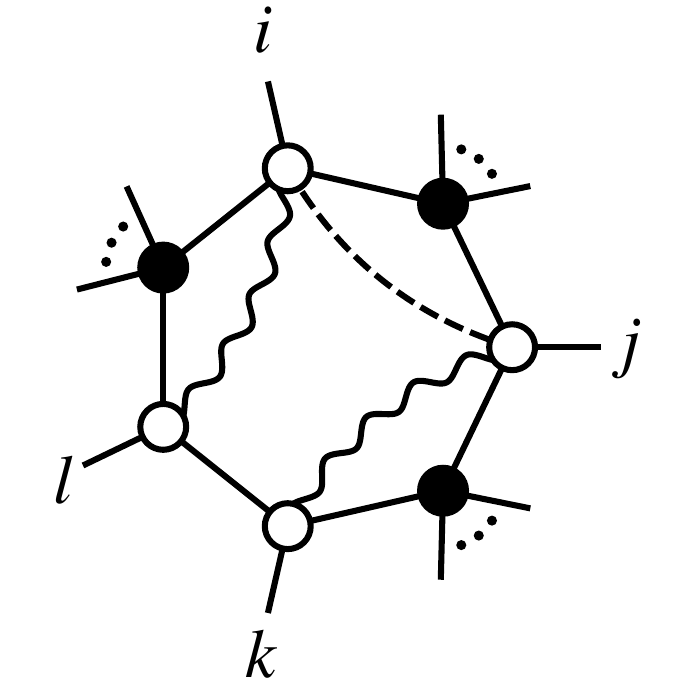}}
    \hspace{-.5cm}
    \raisebox{.4cm}{$\nearrow$}
    \hspace{-.4cm}
    \raisebox{-.4cm}{$\searrow$}
    \hspace{-.4cm}
    \begin{split}
        \includegraphics[scale=.33]{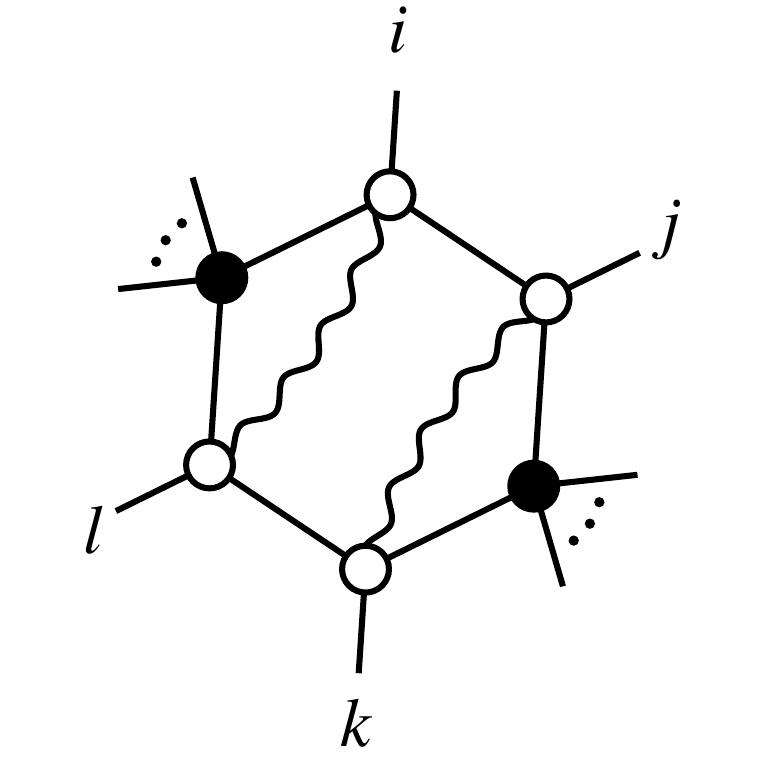} \\[-10pt]
        \includegraphics[scale=.33]{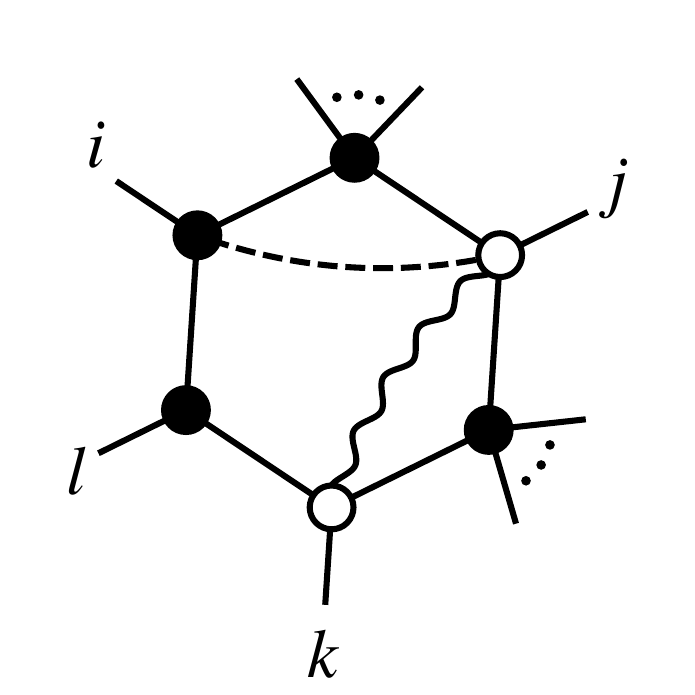}
    \end{split}
    \hspace{-.5cm}
    \raisebox{.4cm}{$\searrow$}
    \hspace{-.4cm}
    \raisebox{-.4cm}{$\nearrow$}
    \hspace{-.6cm}
    \raisebox{-1.05cm}{
    \includegraphics[scale=.33]{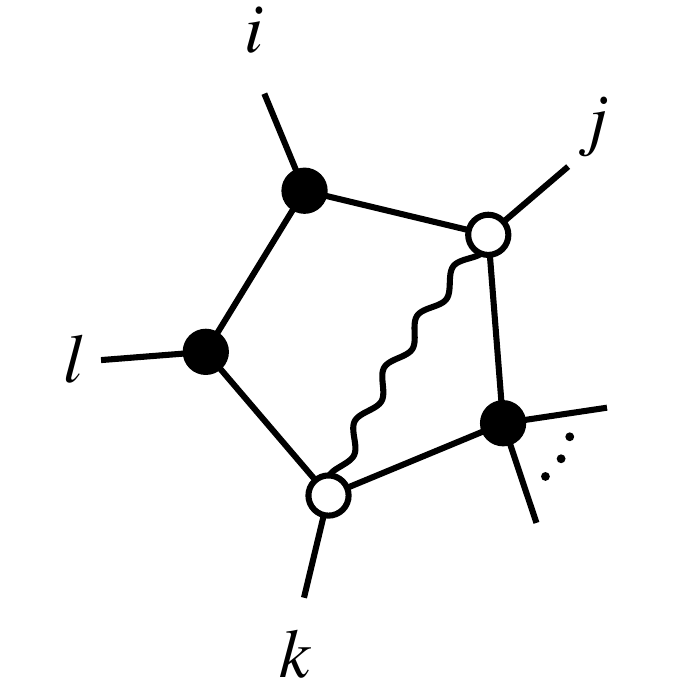}}
\hspace{-1cm}
\end{align}
%
\vskip -.3cm
\noindent
The most degenerate integrals become IR divergent. Additional details regarding these integrands, are discussed in section 5 of \cite{ArkaniHamed:2010gh}, where the interested reader can also find integrated results in terms of dilogarithms. The degenerations of the chiral octagon in eq.~(\ref{eq:octagon_degenerations}) go hand-in-hand with the special sign-flip-four regions in eq.~(\ref{eq:chiral_octagon_sf4_connection}), when pairs of sign flips become adjacent. We have already seen such examples in eq.~(\ref{eq:sf4_chiral_low_n_egs}) and do not display further degenerations for brevity's sake.

If we marginalize over the signs of the $\aab{ik}, \aab{jl}$ brackets that chiralize the sign-flip-four space, we obtain the parity-odd octagon defined in eq.~(\ref{eq:odd_octagon}). (Note that we are \emph{adding} spaces, but the corresponding forms have a relative minus sign which is why marginalizing over the two chiral spaces gives the parity-odd integral.)
\begin{align}
\vspace{-5pt}
  \raisebox{-39pt}{
  \includegraphics[scale=.35]{./figures/sign_flip_four_odd_ijkl}}
  \quad
  \leftrightarrow
  \quad
  \Omega^{\text{o}}_{ijkl}
  \vspace{-15pt}
\end{align}
One aspect that was not emphasized in \cite{ArkaniHamed:2010gh} but is relevant for the connection between these local integrands and the local positive geometries we are discussing here, is a special change of variables that brings the rational integrands, such as the one in eq.~(\ref{eq:chiral_octagon_general}), into $d\log$ form \cite{ArkaniHamed:2012nw,Arkani-Hamed:2014via,Bern:2014kca,Bern:2015ple,Lipstein:2013xra,Henn:2020lye,Brown:2020rda}. This $d\log$ form is not only useful to make a connection to positive geometry, but also leads to simplified differential equations \cite{Herrmann:2019upk}. It turns out that the simplest $d\log$ forms are associated to parity-odd integrands, e.g. 
\begin{align*}
 \Omega^{\text{o}}_{ijkl} {=} 
            d\log\! \frac{\aab{i{-}1i}}{\aab{ii{+}1}}
            d\log\! \frac{\aab{j{-}1j}}{\aab{jj{+}1}}
            d\log\! \frac{\aab{k{-}1k}}{\aab{kk{+}1}}
            d\log\! \frac{\aab{l{-}1l}}{\aab{ll{+}1}}
\end{align*}
Writing compact $d\log$ forms for various integrals is somewhat of an art and for many of the chiral integrals, we do not currently have a simple $d\log$ form available. One notable exception are chiral pentagon integrands, which will play a major role in our companion paper on local triangulations of positive geometries \cite{LocTriangToAppear}.
%
%

%
\vspace{-.5cm}
\section{From dlog Forms to Geometry}
\label{sec:dlog}
\vspace{-.4cm}
%
Thus far, geometric descriptions of scattering amplitudes begin with the definition of a positive geometry, then introduce canonical differential forms with logarithmic singularities on the boundaries of these spaces. This has been true for the Amplituhedron \cite{Arkani-Hamed:2013jha,Arkani-Hamed:2017vfh}, and likewise for the local geometries we have discussed above.

In contrast, in this subsection we turn this story upside-down. There are various Feynman integrals for which $d\log$ forms are known without any a priori connection to some positive geometry. It is natural to ask in what sense these integrals are associated to or give rise to their own geometries. Here, we would like to give a brief appetizer of the types of structures we have uncovered. One-loop box integrals are among the simplest examples where $d\log$ forms are known explicitly. For concreteness, we consider the so-called two-mass-hard configuration  $B_{ii{+}1}$
\vspace{-12pt}
\begin{align}
\label{2mh_boxform}
   &\raisebox{-39pt}{\includegraphics[scale=.6]{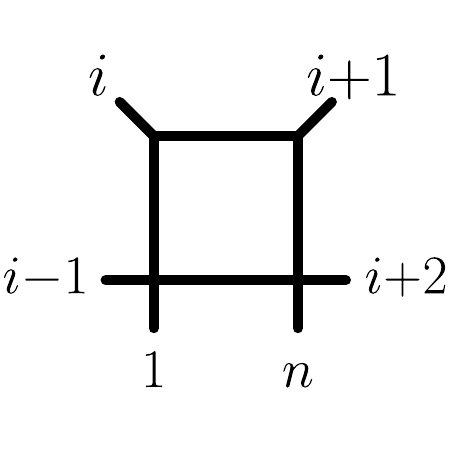}}
   {=} \frac{d^4\mu\, \ab{i{-}1ii{+}1i{+}2}\ab{1ii{+}1n}}
          {\aab{i{-}1i}\aab{ii{+}1}\aab{i{+}1i{+}2}\aab{1n}}
   \\[-10pt]
   &=   d\log\frac{\aab{i{-}1i}}{\aab{X}}
        d\log\frac{\aab{ii{+}1}}{\aab{X}}
        d\log\frac{\aab{i{+}1i{+}2}}{\aab{X}}
        d\log\frac{\aab{1n}}{\aab{X}}\,,
        \nonumber
\end{align}
where $X$ corresponds to either of the two solutions to the quadruple cut $\aab{i{-}1i}{=}\aab{ii{+}1}{=}\aab{i{+}1i{+}2}{=}\aab{1n}{=0}$,
\vspace{-4pt}
\begin{align}
X = \Bigg\{
\begin{split}
   X_i          & =(ii{+}1i{+}2){\cap}(in1)\, \\
   X_{i{+}1}    & =(i{-}1ii{+}1){\cap}(i{+}1n1)\,
\end{split}
\Bigg\}\,.
\vspace{-5pt}
\end{align}
By construction, any space defined by imposing definite signs for the \emph{ratios} of the arguments of each $d\log$ in eq.~(\ref{2mh_boxform}) gives \emph{a} geometry with the correct canonical form. However, simply getting the correct canonical form is insufficient---we also require the exact boundary structure of the geometric space itself. Namely, we demand that $\aab{X}=0$ is not a geometric boundary, and neither are any lower codimension boundaries for which the residues of the form vanish.

This admittedly cryptic statement leads to further constraints that are intricate to implement, and require a slightly more detailed analysis which will be elucidated at length in \cite{LocTriangToAppear}. At the end of the day, we find two consistent geometries originating from the $d\log$ form in eq.~(\ref{2mh_boxform}), both of which have fixed (including negative) signs for the $\aab{ii{+}1}$ brackets of the diagram.
\begin{align}
\vspace{-5pt}
\hspace{-.2cm}
\begin{split}
&
\begin{tabular}{|c||c|c|c|c|c|c|}
\hline \small 
      &  $\!\aab{i{-}1i}\!$       
      & $\!\aab{ii{+}1}\!$ 
      & $\!\aab{i{+}1i{+}2}\!$   
      & $\!\aab{1n}\!$
      & $\!\aab{X_i}\!$          
      & $\!\aab{X_{i{+}1}}\!$ 
    \\[2pt] \hline
    $B^{(1)}_{ii{+1}}$ & $-$ & + & $-$ & + & + & $-$
    \\[2pt] \hline
    $B^{(2)}_{ii{+1}}$ & $-$ & + & $-$ & + & $-$ & $+$
    \\[2pt] \hline
\end{tabular} 
\end{split}
\label{2mh_boxsigns}
\hspace{-.1cm}
\vspace{-5pt}
\end{align}
Both spaces can be considered as two subspaces of a larger achiral space defined only by $\aab{ii{+}1}$ inequalities. The canonical form for this achiral space trivially vanishes because four inequalities are insufficient to produce a non-trivial $d\log$ form with four independent little-group-invariant ratios. As a result, the forms for both $B^{(1)}_{ii{+1}}$ and $B^{(2)}_{ii{+1}}$ are identical up to a sign, and (properly oriented) sum to zero. The achiral space is cut by imposing a single condition on either $\aab{X_i}$ or $\aab{X_{i{+}1}}$, and the respective sign of the other bracket is implied. 

Note that not all signs of $\aab{ii{+}1}$ brackets are fixed in eq.~(\ref{2mh_boxsigns}). Thus, these spaces represent different collections of the sign-flip-four regions introduced above, 
\begin{align}
\label{2mh_box_sf4_expansions}
\begin{split}
    B^{(1)}_{ii{+1}} & \leftrightarrow
    \bigger{
    \sum\limits^{i{-}1}_{l=1}
    \sum\limits^{n}_{k=i{+}2}}
    \hspace{.2cm}
    \raisebox{-39pt}{\includegraphics[scale=.35]{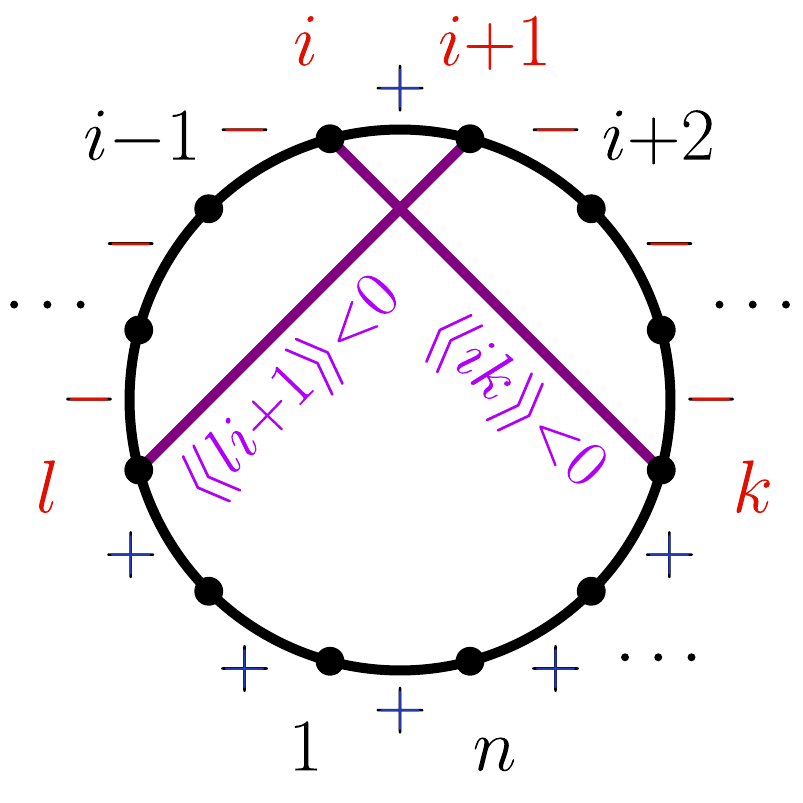}},
\end{split}    
\end{align}
where $B^{(2)}_{ii{+1}}$ is obtained from $B^{(1)}_{ii{+1}}$ by flipping the inequalities of the diagonals $\aab{li{+}1}{<}0\to\aab{li{+}1}{>}0$, and $\aab{ik}{<}0\to\aab{ik}{>}0$. In terms of chirality, the two spaces $B^{(1,2)}_{ii{+1}}$ are parity conjugate, as is apparent from the opposite chirality of all sign-flip-four regions appearing in the expansions of eq.~(\ref{2mh_box_sf4_expansions}) and its $B^{(2)}_{ii{+1}}$ counterpart. 

We can repeat similar analyses for all other integrals entering the `chiral pentagon representation' of one-loop MHV amplitudes in planar $\NeqFour$ \cite{ArkaniHamed:2010kv,ArkaniHamed:2010gh} and associate them to local geometries originating from the study of their $d\log$ forms. As we have seen for the two-mass-hard boxes in eq.~(\ref{2mh_boxsigns}), for individual integrals there can in principle be more than one solution. It is natural to wonder, however, if there is a globally consistent geometry when one attempts to glue individual pieces into a bigger space that describes the full amplitude (without any spurious boundaries). Surprisingly, the result of this exercise selects a unique choice of individual local geometries that combine into a novel positive geometry, which we call the \emph{Amplituhedron-prime}. This space is similar to the original Amplituhedron defined by the inequalities of eq.~(\ref{eq:MHV_A_def}), but is distinct due to the negative signs that appear for various $\aab{ii{+}1}$ brackets, e.g. in both spaces of eq.~(\ref{2mh_boxsigns}). The study of the global consistency of this novel \emph{Amplituhedron-prime} is subtle and we have to defer a detailed discussion to our companion work \cite{LocTriangToAppear}. 

\vspace{-.5cm}
\section{Non-Pure Integrals from non-MHV Kinematics}
\label{sec:nmhv}
\vspace{-.5cm}
%
So far we have focused our discussion on geometric spaces defined for MHV external kinematics where all brackets are positive $\ab{ijkl}>0$ for $i{<}j{<}k{<}l$. In this section, we extend the study of positive geometries beyond the MHV sector. At one loop, the geometry for $\text{N}^k$MHV amplitudes is defined by the sign-flip conditions in eq.~(\ref{eq:AnkL_generic_def}). Although we leave a complete classification for arbitrary $n,k$ to later work, at low multiplicity the picture is simple enough to discuss in this section. At six points, there are three helicity sectors which can be defined by the number of sign flips in the sequence $\{\ab{123i}\}_{i=4,5,6}$, i.e., 
\begin{align}
\label{eq:6pt_ext_data}
\{\underbrace{(+,+,+)}_{\text{MHV}},\underbrace{(+,+,-),(+,-,-)}_{\text{NMHV}},\underbrace{(+,-,+)}_{\text{$\text{N}^2$MHV}}\}.    
\end{align}
The one-loop integrand for arbitrary $k$ is a $4(k{+}1)$ differential form on the space of $\{Z_a,(AB)\}$ with logarithmic singularities on all boundaries. Similarly, for each sign-flip space defined by $\ab{ABii{+}1}$ conditions, the associated form is intimately related to the positivity conditions on the $Z_a$. The positive geometries depend on both the external data and the loop-line and cannot be separated. As such, the logarithmic form is naturally a form in both $dZ_a$ and $d(AB)$. 

We can see this explicitly in the following example: consider the sign-flip-four (in $(AB)$) space
\begin{align}
  \mathcal{S}= \raisebox{-35pt}{\includegraphics[scale=.4]{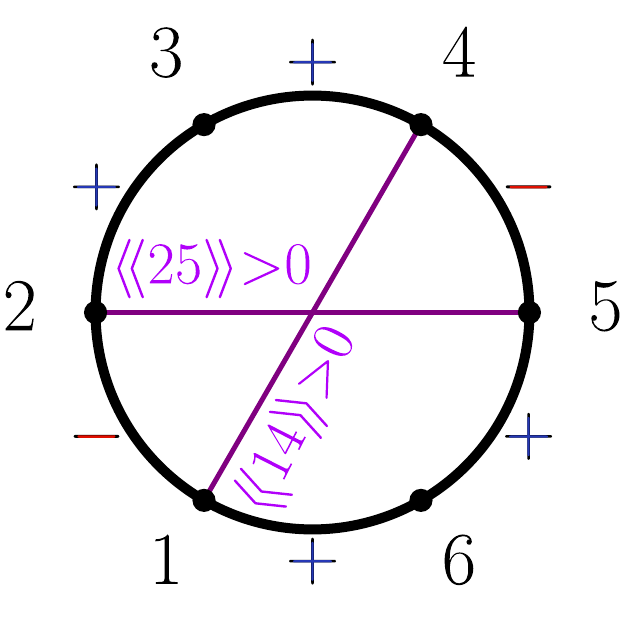}}
 \end{align}
For MHV kinematics the form is the dashed hexagon (which is the parity conjugate of the second example in eq.~(\ref{eq:sf4_chiral_low_n_egs})) and there is no form in $Z_a$. For NMHV kinematics, the calculation is more subtle as the structure of the $(AB)$-space is highly sensitive to the precise signs of brackets $\ab{ijk\ell}$, while the conditions in eq.~(\ref{eq:6pt_ext_data}) are agnostic about the sign of, for example, the four-bracket $\ab{1356}$. In this example, the signs of two four-brackets (not of the form $\ab{ii{+}1jj{+}1}$) are sufficient to fix the associated space in $(AB)$. There are four inequivalent configurations of the external data---each of which has a particular four-form in $Z_a$---that in total yield two distinct local integrands in $(AB)$, which we may label as $\omega^{(1,2)}_{(AB)}$. Labelling the $Z$-space by the signs of $\ab{1235}$ and $\ab{1356}$, respectively, the full eight-form for NMHV kinematics is
\begin{align}
\label{eq:nmhv_form_example}
\hspace{-1cm}
\Omega^{(\mathcal{S})}{=}
\left[\omega^{(++)}_Z{+}\omega^{(--)}_Z\right]\omega^{(1)}_{AB}
{+}
\left[\omega^{(+-)}_Z{+}\omega^{(-+)}_Z\right]\omega^{(2)}_{AB},
\hspace{-.6cm}
\end{align}
where the relevant $Z$-forms, which can be computed along the lines of \cite{Kojima:2020tjf}, are
\begin{align}
\begin{split}
\omega^{(++)}_Z&=\frac{\ab{12356}\ab{13456}^3}{\ab{1235}\ab{1346}\ab{1356}\ab{1456}\ab{3456}},\\
\omega^{(+-)}_Z&=\frac{\ab{13456}\ab{12356}^3}{\ab{1236}\ab{1256}\ab{1345}\ab{1356}\ab{2356}},\\
\omega^{(--)}_Z&=\frac{\ab{12356}\ab{12345}^3}{\ab{1234}\ab{1235}\ab{1245}\ab{1356}\ab{2345}},\\
\omega^{(-+)}_Z&=\frac{\ab{13456}\ab{12345}^3}{\ab{1234}\ab{1245}\ab{1345}\ab{1356}\ab{2345}}.
\end{split}
\end{align}
In the numerators, we use the shorthand notation for e.g.
\begin{align}
\ab{12356}=dZ_1\ab{2356}+\cdots+dZ_6\ab{1235},\,\text{etc.,}
\end{align}
and suppress all wedge products throughout. The two $(AB)$-forms are explicitly
\begin{align}
\begin{split}
\hspace{-.5cm}
\omega^{(1)}_{AB}&{=}\frac{\ab{1235}\left(\ab{1456}\aab{26}\aab{34}{+}\ab{3456}\aab{61}\aab{24}\right)}{\aab{12}\aab{23}\aab{34}\aab{45}\aab{56}\aab{61}},
\hspace{-.4cm}
\\
\hspace{-.5cm}
\omega^{(2)}_{AB}&{=}\frac{\ab{1345}\left(\ab{1256}\aab{23}\aab{46}{-}\ab{1236}\aab{24}\aab{56}\right)}{\aab{12}\aab{23}\aab{34}\aab{45}\aab{56}\aab{61}}.
\hspace{-.4cm}
\end{split}
\end{align}
Nontrivially, in eq.~(\ref{eq:nmhv_form_example}) the spurious poles $\ab{1235}$, $\ab{1356}$, and $\ab{1345}$ cancel, as they were only needed for the triangulation of the full geometry. Note that in the case of $\ab{1235}$, $\ab{1345}$ the cancellations are between the numerators of $\omega_{AB}^{(1,2)}$ and denominators of $\omega_Z^{(\pm,\pm)}$, while $\ab{1356}$ cancels globally between all four pieces in eq.~(\ref{eq:nmhv_form_example}). This shows how inextricably intertwined the $Z_a$ and $(AB)$ parts of $\Omega^{(S)}$ are for higher $k$.

The third helicity configuration relevant for six-point kinematics is $k=2$ i.e. $\text{N}^2$MHV. In this case, for the $(AB)$-conditions $\mathcal{S}$ there is a single $Z$-form,
\begin{align}
\omega_Z^{k{=}2}=\frac{\ab{123456}^4}{\ab{1234}\ab{2345}\ab{3456}\ab{4561}\ab{5612}\ab{6123}},
\end{align}
where the numerator is again shorthand notation e.g. $\ab{123456}^4=(dZ_1)^4(dZ_2)^4\ab{3456}^4+\cdots$. Due to the low multiplicity the associated $(AB)$-form is the same as for the MHV kinematics, eq.~(\ref{eq:sf4_chiral_low_n_egs}), but this does not hold in general. Note that for MHV kinematics the $Z$-form is trivial, $\omega_Z=1$, and only the $(AB)$-part is relevant. 

At higher multiplicities, our preliminary investigations confirm that $(AB)$-spaces with more than four sign flips are \emph{non-empty} for non-MHV helicity configurations. We leave a complete classification for general $n,k$ to future work. These precursory results suggest the non-MHV positive geometries are much richer than their MHV counterparts, not only because of the relevance of $Z$-kinematics, but also because of the presence of higher sign-flip spaces.

\vspace{-.5cm}
\section{Conclusions}
\vspace{-.4cm}

In this letter, we classified the positive geometries which appear in the context of the one-loop Amplituhedron. We discovered that the maximal sign-flip regions are closely connected to chiral octagon integrands, which form the basis of one-loop integrands with special IR properties. This enhances our understanding of positive geometries and their connection to scattering amplitudes, but also opens the door to exploiting the positive geometry framework as a tool to generate infrared finite $d\log$ integrands at two loops and beyond, which would be of eminent importance for modern amplitudes methods.

\vspace{0.2cm}
\textit{Acknowledgment:} This work is supported by DOE grants No. DE-SC0009999, and the funds of University of California. EH is supported by the Department of Energy under contract DE-AC02-76SF00515.

\vspace{-0.5cm}
\bibliographystyle{apsrev4-1} 
\bibliography{amp_refs} 

\end{document}